\documentclass[12pt,preprint]{aastex}

\begin{document}

\title{The Analytical Solution to the Temporal Broadening of a Gaussian-Shaped 
Radio Pulse by Multipath Scattering from a Thin Screen in the Interstellar Medium}
\author{M. M. McKinnon}
\affil{National Radio Astronomy Observatory, Charlottesville, VA \ \ 22903 \ \ USA}

\begin{abstract}
The radio pulse from a pulsar can be temporally broadened by multipath 
scattering in the interstellar medium and by instrumental effects within 
the radio telescope. The observed pulse shape is a convolution of the 
intrinsic one with the impulse responses of the scattering medium and 
instrumentation. Until recently, common methods used to model the observed 
shape make assumptions regarding the intrinsic pulse shape and impulse 
responses, compute the convolution numerically, and solve for the pulse 
width and scattering timescale iteratively. An analytical solution is shown to 
exist for the specific case of the temporal broadening of a Gaussian-shaped pulse 
by a thin scattering screen. The solution is applied to multi-frequency observations 
of PSR B1834--10 to characterize the frequency dependence of its intrinsic pulse 
width and scattering timescale.

\end{abstract}


\section{INTRODUCTION}

The radio pulse from a pulsar can be temporally broadened by multipath 
scattering in the interstellar medium (ISM). The scattered light rays 
traverse a longer path than the rays propagating directly between the 
pulsar and the observer, and the concomitant time delay forms a scattering 
tail on the observed pulse. When the pulsed intensity is averaged over 
many rotations of the star, the observed pulse shape is the convolution 
of the intrinsic pulse shape with the impulse response of the scattering 
medium (Lee \& Jokipii 1975). The analytical form of the impulse
response is determined by the distribution of scattering material 
between the pulsar and the observer (Williamson 1972, 1973). If the 
scattering material is concentrated in a region that is thin in comparison 
to the distance between the pulsar and the observer and the plasma 
inhomogeneities in it follow a Gaussian power spectrum, the impulse 
response is a truncated exponential characterized by a scattering timescale, 
$\tau_s$ (Cronyn 1970; Lee \& Jokipii 1975). Many pulse shapes observed to 
have a scattering component are consistent with this \lq\lq thin screen" 
approximation (Komesaroff et al. 1972; L\"ohmer et al.  2001, 2004; Bhat et al. 2004).

Free electrons in the ISM cause the broad-band radio pulse to be dispersed 
in time, with high frequency pulses arriving earlier than low frequency 
ones. The frequency dependence of the pulse arrival times follows a cold 
plasma dispersion law. The magnitude of the effect is characterized by the 
dispersion measure (DM), which is the electron density integrated along 
the line of sight to the pulsar. Since DM can be accurately measured, and 
only varies slowly with time if at all, its effects can be removed for the 
most part in constructing pulsar average profiles. Pulsar DMs and scattering 
timescales, which vary with both DM and frequency (e.g. Bhat et al. 2004), 
tend to be larger for those pulsars within the Galactic Plane and at large 
distances. DM measurements and independent estimates of pulsar distances can 
be used to construct models of the distribution of free electrons in the Galaxy 
(Taylor \& Cordes 1993; Cordes \& Lazio 2002). The models can be extended 
to include interstellar scattering using the scattering timescales extracted 
from pulsar average profiles (e.g. Cordes, Weisberg, \& Boriakoff 1985; 
Bhat et al. 2004). 

Pulse profiles computed from multi-frequency observations can be used to 
study the frequency dependence of $\tau_s$ and the intrinsic width of the 
pulsar's pulse. From  a theoretical perspective, the scattering timescale 
should follow a power law, $\tau_s\propto\nu^{-\alpha}$, where the frequency
scaling index is $\alpha=4$ if the plasma inhomogeneities in the scattering 
medium follow a Gaussian power spectrum and $\alpha=4.4$ for a Kolmogorov 
spectrum (Lee \& Jokipii 1975; Rickett 1977). Multi-frequency observations 
of pulsars indicate the measured values of $\alpha$ are generally consistent 
with the theoretical ones (Komesaroff et al.  1972), although smaller scaling
indices have been reported for highly dispersed pulsars (L\"ohmer et al. 2004; 
Bhat et al. 2004), suggesting a more uniform distribution of the scattering 
material and that the plasma inhomogeneties are non-Gaussian, perhaps following 
L\'evy-Cauchy statistics (Boldyrev \& Gwinn 2005). The frequency dependence of 
the intrinsic pulse width is also modeled as a power law with a scaling index 
in the range of 0.14 to 0.45, depending upon the model assumptions (Thorsett 
1991 and references therein). The model of Ruderman \& Sutherland (1975), for 
example, assumes the magnetic field in the emission region is dipolar in shape, 
the radiation is emitted at the local plasma frequency that varies with height 
in the magnetosphere, and the angular extent of the radiation is determined by 
the last field line to close within the star's corotating magnetosphere. These 
assumptions have led to the popular concept of a radius-to-frequency 
mapping (Cordes 1978) with the pulse width varying with frequency as 
$\nu^{-1/3}$. In general, pulse profiles are observed to widen with 
decreasing frequency. Observed values of the width's frequency scaling index,
however, lie in the range of about 0 to 1, and the width can tend towards 
a constant value at high frequency (Rankin 1983; Thorsett 1991). Owing to 
the different scaling indices in the frequency dependencies of the 
intrinsic width and scattering timescale, the observed pulse shape is 
heavily influenced by multipath scattering at low radio frequencies 
($\nu\sim 100$ MHz), while it is essentially a replica of the intrinsic 
shape at high frequencies ($\nu\sim 10$ GHz). At moderate frequencies 
($\nu\sim 1$ GHz), the pulsar's scattering timescale and intrinsic width 
can have comparable weight in shaping the observed pulse waveform.

Instrument related effects can also broaden the radio pulse in time.
Wide bandwidth observations of pulsars are separated into narrow frequency 
channels to compensate for the effects of interstellar dispersion. The 
dispersion smearing of the pulse over this narrow detection bandwidth leads 
to a dispersion impulse response $d(t)$ characterized by a frequency-dependent 
timescale, $\tau_d$. Additionally, detection circuitry in the radio 
telescope does not necessarily respond instantaneously to the pulse. It, 
too, has an impulse response, $i(t)$, with a time constant, $\tau_i$. The 
observed pulse, $f(t)$, is then a convolution of the intrinsic pulse, $g(t)$, 
with the impulse responses of both the scattering medium, $s(t)$, and the 
instrument.

\begin{equation}
f(t) = i(t)*d(t)*s(t)*g(t)
\label{eqn:obspulse}
\end{equation}

A variety of methods have been developed to determine the intrinsic pulse 
width and the scattering timescale from the observed pulse profile by 
solving Equation~\ref{eqn:obspulse}. All have done so by computing the 
convolution numerically on an iterative basis and assuming functional 
forms for the instrumental impulse response functions. The first method 
adopts the thin screen approximation for $s(t)$, assumes the intrinsic pulse 
shape is a Gaussian, or sum of Gaussians, computes the convolution given by 
Equation~\ref{eqn:obspulse} numerically, and solves for the pulse width and 
scattering timescale iteratively with a least squares fit of the observed 
profile to the computed profile (e.g. Ramachandran et al. 1997; Mitra \& 
Ramachandran 2001; L\"ohmer et al. 2001, 2004). A second method uses Fourier 
inversion to deconvolve the intrinsic pulse shape from the impulse responses 
of the instrument and ISM. Since the Fourier transform of a convolution of 
functions is the product of their Fourier transforms (Bracewell 1986), one 
can compute the intrinsic pulse profile by assuming an $s(t)$, dividing the 
Fourier transform of the observed profile by the Fourier transform of $s(t)$, 
and Fourier transforming the result back to the time domain. Kuzmin \& 
Izvekova (1993) demonstrated this method using different scattering functions. 
Bhat, Cordes, \& Chatterjee (2003) developed a third method that makes no 
assumptions regarding the specific functional form of the intrinsic pulse 
shape and uses a CLEAN-based algorithm to determine the scattering timescale 
iteratively by requiring the deconvolved intrinsic profile to have \lq\lq 
minimum asymmetry" and non-negative values of total intensity. The method 
also adopts the thin screen approximation for $s(t)$. Bhat et al. (2004) 
extended this method to include a scattering function that corresponds to a 
uniformly distributed scattering medium with a square-law structure function. 
More recently, Demorest (2010) demonstrated a fourth method for determining 
the scattering function and intrinsic pulse profile that is based upon cyclic 
spectral analysis. Traditional spectral analysis procedures assume the pulsar 
signal is stationary over the timescale of the observation and only utilize 
the signal's amplitude. Demorest recognized that the pulsar signal is more 
accurately described as cyclo-stationary, allowing one to use the signal's 
amplitude and phase to determine $s(t)$ and $g(t)$ without any assumptions 
regarding their functional forms. Applying the method to PSR B1937+21, 
Demorest found the shapes of its pulse components to be Gaussian-like, while 
$s(t)$ was comprised of an exponential decay followed by a slowly decaying 
tail. As undeniably powerful as this method is, it can be computationally 
expensive, and recent work, such as that for fast radio bursts (Thornton et 
al. 2013), continues to use the more traditonal methods described above to 
estimate scattering timescales.

The purpose of this paper is to show that simple analytical solutions 
to Equation~\ref{eqn:obspulse} exist in the specific case where the intrinsic 
pulse shape is Gaussian and $s(t)$ is the truncated exponential for a thin 
scattering screen. The solutions are derived in Section~\ref{sec:analysis} 
using functional forms of $i(t)$ and $d(t)$ that are commonly found in the 
literature. In Section~\ref{sec:apply}, the solution is applied to 
multi-frequency observations of PSR B1843--10 to illustrate the applicability 
of the solutions and to investigate the frequency dependence of the pulsar's 
width and scattering timescale.

\section{TEMPORAL BROADENING ANALYSIS}
\label{sec:analysis}

The pulse broadening problem posed by the multiple convolutions in
Equation~\ref{eqn:obspulse} has an analytical solution when $g(t)$ is 
Gaussian, $s(t)$ is a truncated exponential, and $i(t)$ and $d(t)$ are 
any combination of Gaussians, truncated exponentials, and delta functions. 
The Gaussian pulse shape is 

\begin{equation}
g(t,\mu) = {S\over{\sigma\sqrt{2\pi}}}
     \exp{\Biggl[-{1\over{2}}\Biggl({t-\mu\over{\sigma}}\Biggr)^2\Biggr]}, 
\label{eqn:gauss}
\end{equation}
where $S$ is the integrated flux density of the pulse, $\mu$ is its 
centroid, and $\sigma$ is its standard deviation. The truncated 
exponential is

\begin{equation}
e(t,\tau) = {1\over{\tau}}\exp(-t/\tau)H(t),
\label{eqn:exp}
\end{equation}
where $\tau$ is a time constant and $H(t)$ is the unit step, or Heaviside, 
function. $H(t)$ is equal to zero for $t<0$, and one otherwise. Scattering 
conserves energy (Cronyn 1970), and the factor of $1/\tau$ in 
Equation~\ref{eqn:exp} normalizes the impulse response so that the 
integrated flux density of the radio pulse is conserved in the convolution 
(Bhat, Cordes, \& Chatterjee 2003). 

\subsection{Broadening Due to Scattering Only}

In the absence of instrumental effects (i.e. $i(t)=d(t)=\delta(t)$), the pulse
is broadened only by multipath scattering in the ISM, and the shape of the
observed pulse is given by the convolution of Equations~\ref{eqn:gauss} 
and~\ref{eqn:exp}.

\begin{equation}
f(t,\tau) = {S\over{2\tau}}\exp{\Biggl({\sigma^2\over{2\tau^2}}\Biggr)} 
     \exp{\Biggl[-{(t-\mu)\over{\tau}}\Biggr]} 
     \Biggl\{1 + {\rm erf} \Biggl[{t-(\mu +\sigma^2/\tau)
     \over{\sigma\sqrt{2}}}\Biggr]\Biggr\}
\label{eqn:tempbroad}
\end{equation}

\noindent The function $ w(x)=[1+{\rm erf}(x)]/2$ within 
Equation~\ref{eqn:tempbroad} varies antisymmetrically from zero to one about 
$t=\mu + \sigma^2/\tau$, where its slope is maximum and varies inversely with 
$\sigma$. It attenuates the waveform for times much less than 
$t=\mu + \sigma^2/\tau$, and evolves into the step function, $H(t-\mu)$, as 
the pulse width becomes small with respect to the scattering timescale, 
$\sigma/\tau\ll 1$. In this case, $g(t)$ is a delta function in comparison to 
$s(t)$, and the convolution of $g(t)$ with $s(t)$ essentially replicates the 
exponential at $t=\mu$. Conversely, $f(t,\tau)$ is Gaussian-like in shape when 
$\sigma/\tau \gg 1$. 

\subsection{Broadening with Instrumental Effects Included}

The instrumental impulse responses $i(t)$ and $d(t)$ in 
Equation~\ref{eqn:obspulse} are assumed to have a variety of functional 
forms. For example, Ramachandran et al. (1997) and Mitra \& Ramachandran (2001) 
consider the dispersion smearing and instrumental responses to be negligible. 
Thus, they assume the instrumental impulse responses are delta functions
($i(t)=d(t)=\delta(t)$), and Equation~\ref{eqn:tempbroad} is the analytical 
form of the numerical model they adopted. Similarly, L\"ohmer et al. (2001, 2004) 
suggest the rise times of the telescope receivers and back ends are small enough 
to consider the effect of $i(t)$ negligible ($i(t)=\delta(t)$). They model $d(t)$ 
as a rectangular function of width $\tau_d$ for incoherent dedispersion and 
$d(t) = \delta(t)$ for coherent dedispersion. Therefore, 
Equation~\ref{eqn:tempbroad} is the analytical form of their numerical model 
for their coherently dedispersed data.

Bhat, Chatterjee, \& Cordes (2003) suggest the term $i(t)*d(t)$ is itself 
a convolution of many different truncated exponentials and effectively invoke 
the central limit theorem to argue that its functional form is Gaussian. Since 
the convolution of two or more Gaussians is a Gaussian (e.g. Bracewell 1986), 
the broadening problem becomes the convolution of a single Gaussian with the 
truncated exponential from multipath scattering in the ISM. 
Equation~\ref{eqn:tempbroad} represents the analytical form of the numerical 
model adopted by Bhat, Chatterjee, \& Cordes for the specific case when the 
intrinsic shape of their minimum asymmetry waveform is Gaussian, with $\mu$ 
in the equation replaced by the sum of the means of all Gaussian waveforms 
and $\sigma$ replaced by the root sum square of their standard deviations. 
The equation also represents the solution to the pulse broadening problem when 
the instrumental impulse responses are both Gaussian in shape. 

Cordes, Weisberg, \& Boriakoff (1985) assume the instrumental impulse responses
are truncated exponentials with different time constants. Post-detection signal 
integration in an analog filter bank can be accomplished with a simple 
electronic circuit consisting of a resistor and capacitor wired in series. 
The impulse response of the circuit is a truncated exponential with a 
post-detection time constant given by the product of the resistance and 
capacitance (e.g. Nayfeh \& Brussel 1985; Fink \& Christiansen 1982). 
When the impulse responses of the instrument and ISM are truncated 
exponentials, the observed waveform is 

\begin{equation}
h(t) = e(t,\tau_i)*e(t,\tau_d)*e(t,\tau_s)*g(t) = E(t)*g(t),
\end{equation}
where $e(t,\tau)$ is given by Equation~\ref{eqn:exp}, $\tau_s,\tau_d$ 
and $\tau_i$ are the time constants for each impulse response, and $E(t)$ 
defines an impulse response for the entire system. Multiple convolutions 
of truncated exponentials form a linear combination of truncated 
exponentials (cf Williamson 1973), so that the system impulse response is

\begin{equation}
E(t) = \Biggl[{\tau_s\exp(-t/\tau_s)\over{(\tau_s-\tau_d)(\tau_s-\tau_i)}}
     - {\tau_d\exp(-t/\tau_d)\over{(\tau_s-\tau_d)(\tau_d-\tau_i)}} 
     + {\tau_i\exp(-t/\tau_i)\over{(\tau_s-\tau_i)(\tau_d-\tau_i)}}
       \Biggr]H(t).
\label{eqn:SIR}
\end{equation}
The observed pulse, $h(t)$, is simply a linear combination of 
Equation~\ref{eqn:tempbroad} with different time constants.

\begin{equation}
h(t) = {\tau_s^2f(t,\tau_s)\over{(\tau_s-\tau_i)(\tau_s-\tau_d)}} 
     - {\tau_i^2f(t,\tau_i)\over{(\tau_s-\tau_i)(\tau_i-\tau_d)}}
     + {\tau_d^2f(t,\tau_d)\over{(\tau_s-\tau_d)(\tau_i-\tau_d)}}.
\label{eqn:general}
\end{equation}
Values of $h(t)$ calculated with Equation~\ref{eqn:general} are always 
non-negative, regardless of the time constant values. The equation is 
written under the assumption that $\tau_s>\tau_i>\tau_d$ so that the 
coefficients for each of the three $f(t,\tau)$ terms in it are 
non-negative. If the pulse width is also less than the scattering 
timescale but greater than the dispersion smearing and instrument 
response times, as is generally the case, the leading term in 
Equation~\ref{eqn:general} has the classic shape of an 
exponentially-broadened pulse, and the two trailing terms have narrow 
Gaussian shapes. Equation~\ref{eqn:general} is the analytical form of 
the numerical model adopted by Cordes, Weisberg, \& Boriakoff (1985) 
when the intrinsic pulse shape is Gaussian (see, for example, their
Figure 1).

\begin{figure}
\plotone{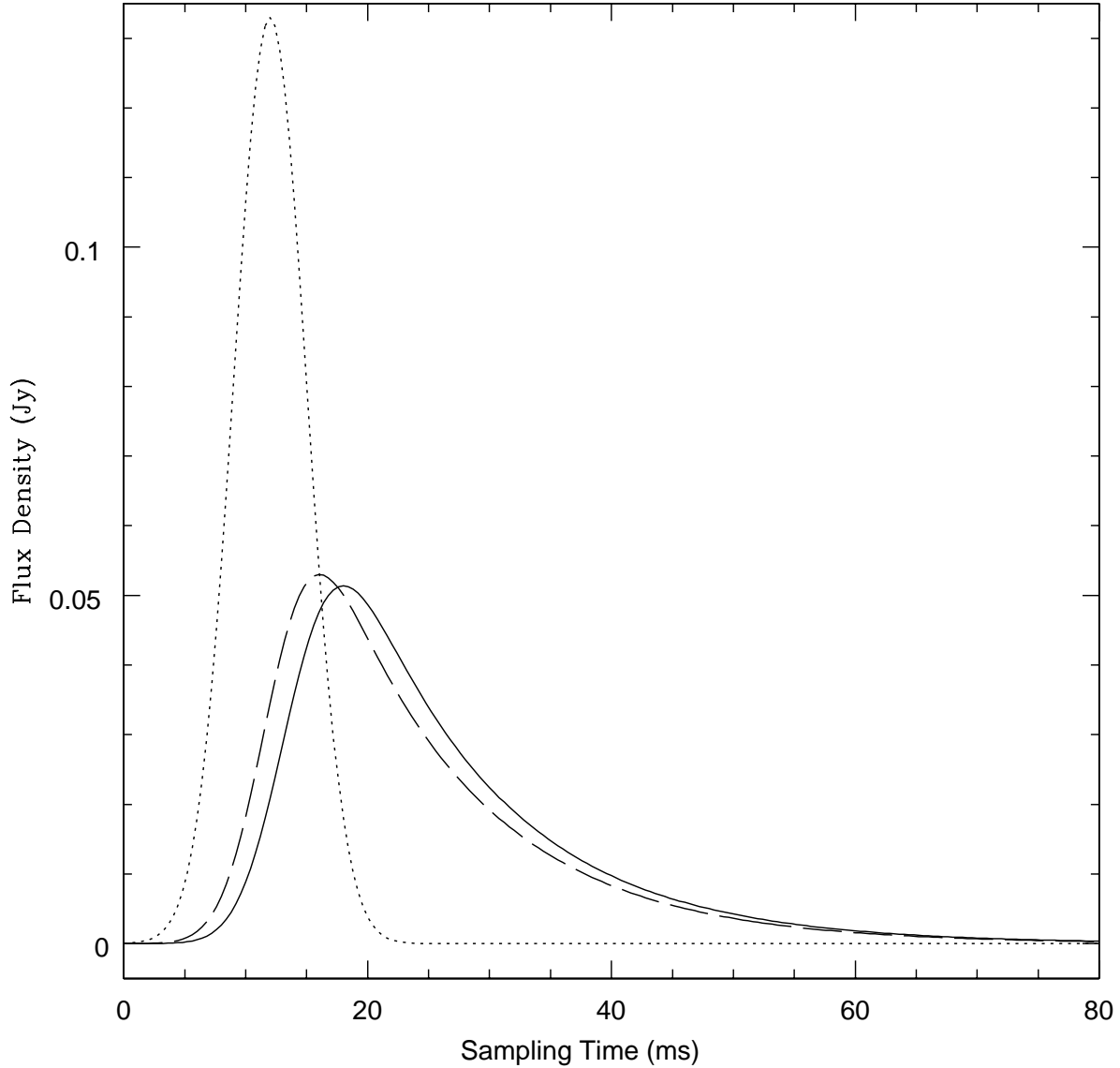}
\caption{Temporal broadening of a Gaussian-shaped pulse. The pulsar-intrinsic 
Gaussian pulse (Eqn.~\ref{eqn:gauss}) is shown by the dotted line. The 
intrinsic pulse broadened by scattering from a thin screen in the ISM 
(Eqn.~\ref{eqn:tempbroad}), exclusive of any instrumental effects, is shown 
by the dashed line. The observed pulse with instrumental broadening terms 
included (Eqn.~\ref{eqn:general}) is shown by the solid line. The parameters 
used to calculate the waveforms in the figure are $S=1$ Jy,  $\mu=12$ ms, 
$\sigma=3$ ms, $\tau_s = 12$ ms, $\tau_i=1$ ms, and $\tau_d=0.8$ ms.}
\label{fig:tempbroad}
\end{figure}

Examples of pulsar-intrinsic and scattered waveforms are shown in 
Figure~\ref{fig:tempbroad}. The Gaussian profile of the intrinsic pulse 
(Eqn.~\ref{eqn:gauss}) is shown by the dotted line. The dashed line shows 
the intrinsic pulse scattered by a thin screen in the ISM 
(Eqn.~\ref{eqn:tempbroad}), exclusive of any instrumental effects. The 
solid line shows the final observed pulse with instrumental effects 
included, assuming the instrumental impulse responses are truncated 
exponentials as included in Equation~\ref{eqn:general}. The parameters 
used to compute the waveforms are listed in the figure caption. The 
integrated flux densities of the three waveforms are identical. 

As written, the equations for the system impulse response
(Eqn.~\ref{eqn:SIR}) and the observed pulse (Eqn.~\ref{eqn:general}) 
are indeterminate when any two, or all three, time constants are equal. 
When two of the time constants are equal, say $\tau_d=\tau_i$, the 
system impulse response and observed pulse are represented by 

\begin{equation}
E(t) = {\tau_s\over{(\tau_s-\tau_d)^2}}
   \Biggl\{\exp{\Biggl(-{t\over{\tau_s}}\Biggr)} - \exp{\Biggl(-
   {t\over{\tau_d}}\Biggr)}\Biggl[1 + {t(\tau_s-\tau_d)\over{\tau_s\tau_d}}
    \Biggr]\Biggr\}H(t),
\end{equation}

\begin{eqnarray}
h(t) & = & \Biggl({\tau_s\over{\tau_s-\tau_d}}\Biggr)^2f(t,\tau_s)
     - \Biggl[{\tau_s\tau_d\over{(\tau_s-\tau_d)^2}} 
  + {t-(\mu+\sigma^2/\tau_d)\over {(\tau_s-\tau_d)}}\Biggr]f(t,\tau_d) 
   \nonumber \\
   & - & {\sigma^2\over{\tau_d(\tau_s-\tau_d)}}
   \exp{\Biggl({\sigma^2\over{2\tau_d^2}}\Biggr)}
    \exp{\Biggl[-{(t-\mu)\over{\tau_d}}\Biggr]}g(t,\mu+\sigma^2/\tau_d),
\end{eqnarray}
where $f(t,\tau_s)$, $f(t,\tau_d)$ are given by Equation~\ref{eqn:tempbroad}
and $g(t,\mu+\sigma^2/\tau_d)$ is given by Equation~\ref{eqn:gauss}.  When 
all three time constants are equal to $\tau$, the system impulse response 
and the observed pulse are

\begin{equation}
E(t) = \Biggl({t\over{\tau}}\Biggr)^2{e(t,\tau)\over{2}},
\end{equation}

\begin{eqnarray}
h(t) & = & {[t-(\mu + \sigma^2/\tau)]^2 + \sigma^2\over{2\tau^2}}f(t,\tau) 
            \nonumber \\
     & + & \Biggl({\sigma\over{\tau}}\Biggr)^2\exp{\Biggl( 
    {\sigma^2\over{2\tau^2}}\Biggr)}\exp{\Biggl[-{(t-\mu)\over{\tau}}
    \Biggr]}{t-(\mu + \sigma^2/\tau)\over{2\tau}}g(t,\mu + \sigma^2/\tau).
\end{eqnarray}

In another instrumental configuration where $i(t)$ is a delta
function and $d(t)$ is a truncated exponential, the system impulse
response and observed pulse are given by

\begin{equation}
E(t) = {\exp(-t/\tau_s)-\exp(-t/\tau_d)\over{\tau_s-\tau_d}}H(t),
\label{eqn:SIRdelta}
\end{equation}

\begin{equation}
h(t) = {\tau_sf(t,\tau_s) -  \tau_df(t,\tau_d)
       \over{\tau_s-\tau_d}}.
\label{eqn:idelta}
\end{equation}
Equation~\ref{eqn:SIRdelta} is also the mathematical representation of the 
two-component, thin screen scattering model discussed by Rankin \& 
Counselman (1973), Isaacman \& Rankin (1977), and Williamson (1973). 
Additionally, the equation may be interpreted as the representation of a 
pulse with the intrinsic shape of a truncated exponential that has been 
scattered by a single thin screen in the ISM. Equations~\ref{eqn:SIRdelta} 
and~\ref{eqn:idelta} are also indeterminate when $\tau_s=\tau_d$. In this 
case, the system impulse response and observed pulse are given by

\begin{equation}
E(t) = {t\over{\tau}}e(t,\tau),
\end{equation}

\begin{equation}
h(t) = {t-(\mu + \sigma^2/\tau)\over{\tau}}f(t,\tau) 
     +  \Biggl({\sigma\over{\tau}}\Biggr)^2\exp{\Biggl( 
    {\sigma^2\over{2\tau^2}}\Biggr)}\exp{\Biggl[-{(t-\mu)\over{\tau}}
    \Biggr]}g(t,\mu + \sigma^2/\tau).
\end{equation}

\section{APPLICATION}
\label{sec:apply}

Multi-frequency observations of PSR B1834--10 were used to characterize the 
frequency dependence of the pulsar's scattering timescale and intrinsic pulse 
width by least squares fits of the data to relevant expressions derived in 
Section~\ref{sec:analysis}. The data were recorded by Gould \& Lyne (1998) and 
are publicly available through the European Pulsar Network. The dispersion measure 
and topocentric rotation period of the pulsar at the time of the observation were
318.0 ${\rm pc\ cm}^{-3}$ and 563.707 ms, respectively. The instrumentation used to 
record the data consisted of a conventional analog filter bank followed by square 
law detectors. Therefore, Equation~\ref{eqn:general} is generally an appropriate 
model for the observation and was used in the fit to the data.

The parameters of the observations are described in Table 1. The table lists the 
observing frequency, the total bandwidth of the observation, the filter bandwidth, 
the number of samples ($N$) recorded across the pulsar's period, the sampling 
interval ($t_s$), and the dispersion smearing time over the filter bandwidth 
($\tau_d$). Gould \& Lyne did not report the values of their post-detection 
integration times ($\tau_i$), so $\tau_i$ was assumed to equal $t_s$ in the fits.
They also observed the pulsar at 408 MHz; however, that observation was not used 
in the analysis because its temporal resolution was too coarse ($N=58$).

\begin{figure}
\plotone{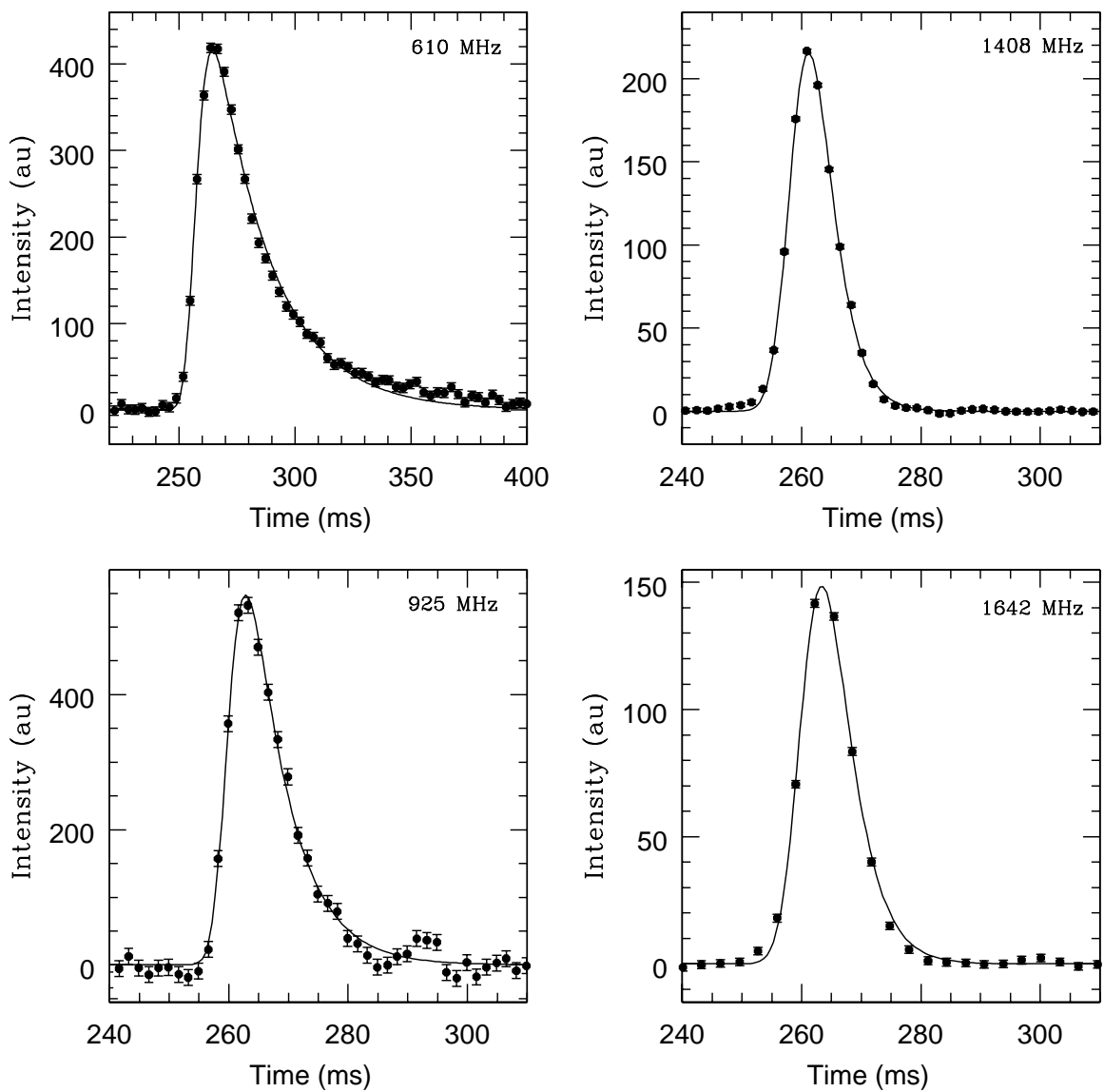}
\caption{Temporal broadening of the pulse from PSR B1834--10 at four frequencies. 
The solid line in each panel of the figure is the best fit of 
Equation~\ref{eqn:general} to the data. The fit parameters are listed in Table 2. 
The scale of the error bars on the data points is equal to the off-pulse 
instrumental noise. The units of intensity are arbitrary (au).}
\label{fig:profiles}
\end{figure}

The results of the fits are listed in Table 2 and are illustrated in 
Figure~\ref{fig:profiles}. The pulse profiles at all four frequencies are simple, 
consisting of a single component, with the possible exception of an additional 
component of low intensity appearing on the trailing edge of the 925 MHz profile. 
The figure shows the model is a good representation of all four data sets, although 
the model slightly underestimates the 610 MHz profile in the region of the scattering 
tail. There is no evidence for temporal broadening due to interstellar scattering at 
1642 MHz as that profile is consistent with a single component broadened only by 
instrumental effects. The large dispersion smearing caused by the large filter 
bandwidth used in the 1642 MHz observation causes the observed profile at that 
frequency to be wider than what one might expect from the profiles at 925 MHz 
and 1408 MHz.

A linear regression of the scattering timescales determined from the fits shows 
they follow a power law with a scattering timescale at 1 GHz of 
$\tau=6.3^{+3.1}_{-2.1}$ ms and a frequency scaling index of $\alpha=2.5^{+1.2}_{-1.1}$. 
Similar to results in L\"ohmer et al. (2001) and Bhat et al. (2004), the scaling  
index is smaller than the values of 4.0 or 4.4 expected from a Gaussian or Kolmogorov 
power spectrum for plasma inhomogeneities in the scattering medium. This result 
is also inconsistent with the assumption adopted in the derivation that the 
scattering is concentrated in a thin screen. The intrinsic widths, as represented 
by the values of $\sigma$ determined from the fits, do not vary systematically with 
frequency, and are consistent within the errors with a constant value of 
$\sigma=2.5$ over the observed frequency range. 

\section{SUMMARY}

The temporal broadening of a Gaussian-shaped pulse by multipath scattering 
from a thin screen in the ISM was derived analytically for a variety of 
instrumental impulse response functions for the first time. The result was 
used to characterize the frequency dependence of the intrinsic pulse width 
and scattering timescale of PSR B1834--10. The intrinsic shape of the pulse 
is consistent with a single Gaussian component that varies little in width 
over a frequency range of 610 MHz to 1642 MHz.  The frequency dependence of 
the pulsar's scattering timescale is flatter than what is expected from the 
canonical Gaussian and Kolmogorov power spectra for plasma inhomogeneities 
in the scattering medium.  

\acknowledgements The National Radio Astronomy Observatory is a facility of the 
National Science Foundation operated under cooperative agreement by Associated 
Universities, Inc.  Part of this research has made use of the data base of
published pulse profiles maintained by the European Pulsar Network, available at 
http://www.mpfir-bonn.mpg.de/pulsar/data/.

\begin{deluxetable}{cccccc}
\tablenum{1}
\tablewidth{320pt}
\tablecaption{Observational Parameters}
\tablehead{
\colhead{Frequency} & \colhead{Total BW} & \colhead{Filter BW} & \colhead{$N$} & 
\colhead{$t_s$} & \colhead{$\tau_d$} \\
\colhead{(MHz)} & \colhead{(MHz)} & \colhead{(MHz)} & \colhead{} &
\colhead{(ms)} & \colhead{(ms)}}
\startdata
 610 & 4.0 & 0.125 & 189 & 3.0 & 1.5 \\
 925 & 8.0 & 0.250 & 337 & 1.7 & 0.8 \\
 1408 & 32.0 & 1.000 & 301 & 1.8 & 1.0 \\
 1642 & 40.0 & 5.000 & 177 & 3.2 & 3.0 \\
\enddata
\end{deluxetable}

\begin{deluxetable}{cccccc}
\tablenum{2}
\tablewidth{360pt}
\tablecaption{Results of Least Squares Fits}
\tablehead{
\colhead{Frequency} & \colhead{$S$} & \colhead{$\mu$} & \colhead{$\tau_s$} &
\colhead{$\sigma$} & \colhead{$\chi^2$} \\
\colhead{(MHz)} & \colhead{(au)} & \colhead{(ms)} & \colhead{(ms)} & 
\colhead{(ms)} & \colhead{}}
\startdata
 610 & $5000\pm 260$ & $250.7\pm 0.8$ & $24.4\pm 2.0$ & $3.2\pm 1.2$ & 1.63 \\
 925 & $3910\pm 480$ & $256.4\pm 0.7$ & $6.4\pm 1.4$ & $1.7\pm 1.1$ & 1.11 \\
 1408 & $1110\pm 50$ & $254.9\pm 0.2$ & $2.9\pm 0.3$ & $2.4\pm 0.3$ & 1.62 \\
 1642 & $520\pm 30$  & $255.9\pm 0.4$ &    --        & $2.6\pm 0.4$ & 1.13 \\
\enddata
\end{deluxetable}


\begin{references}

\reference{} Bhat, N. D. R., Cordes, J. M., \& Chatterjee, S., 2003,
             \apj, 584, 782

\reference{} Bhat, N. D. R., Cordes, J. M., Camilo, F., Nice, D. J., \& 
             Lorimer, D. R., 2004, \apj, 605, 759

\reference{} Boldyrev, S. \& Gwinn, C. R., 2005, \apj, 624, 213

\reference{} Bracewell, R. N., 1986, The Fourier Transform and Its 
             Applications, New York: McGraw-Hill, 230

\reference{} Cordes, J. M., 1978, \apj, 222, 1006

\reference{} Cordes, J. M., Weisberg, J. M., \& Boriakoff, V., 1985, \apj, 
             288, 221

\reference{} Cordes, J. M. \& Lazio, T. J. W., 2002, astro-ph/0207156v3

\reference{} Cronyn, W. M., 1970, Science, 168, 1453

\reference{} Demorest, P. B., 2011, \mnras, 416, 2821

\reference{} Fink, D. G. \& Christiansen, D., 1982, Electronics Engineers'
             Handbook, New York: McGraw-Hill, 16-2

\reference{} Gould, D. M. \& Lyne, A. G., 1998, \mnras, 301, 235

\reference{} Isaacman, R. \& Rankin, J. M., 1977, \apj, 214, 214

\reference{} Komesaroff, M. M., Hamilton, P. A., \& Ables, J. G., 1972,
             Aust. J. Phys., 25, 759

\reference{} Kuzmin, A. D. \& Izvekova, V. A., 1993, \mnras, 260, 724

\reference{} Lee, L. C. \& Jokipii, J. R., 1975, \apj, 201, 532.

\reference{} L\"ohmer, O., Kramer, M., Mitra, D., Lorimer, D. R., \& Lyne,
             A. G., 2001, \apj, 562, L157

\reference{} L\"ohmer, O., Mitra, D., Gupta, Y., Kramer, M., \& Ahuja, A.,
             2004, A\&A, 425, 569

\reference{} Mitra, D. \& Ramachandran, R., 2001, A\&A, 370, 586 

\reference{} Nayfeh, M. H. \& Brussel, M. K., 1985, Electricity and Magnetism, 
             New York: Wiley, 397 

\reference{} Ramachandran, R., Mitra, D. Deshpande, A. A., McConnell, D. M., 
             \& Ables, J. G., 1997, \mnras, 290, 260 

\reference{} Rankin, J. M., 1983, \apj, 274, 359 

\reference{} Rankin, J. M. \& Counselman, C. C., 1973, \apj, 181, 875 

\reference{} Rickett, B. J., 1977, \araa, 15, 479

\reference{} Ruderman, M. A. \& Sutherland, P. G., 1975, \apj, 196, 51

\reference{} Taylor, J. H. \& Cordes, J. M., 1993, \apj, 411, 674

\reference{} Thorsett, S. E., 1991, \apj, 377, 263

\reference{} Thornton, D. et al., 2013, Science, 341, 53

\reference{} Williamson, I. P., 1972, MNRAS, 157, 55

\reference{} Williamson, I. P., 1973, MNRAS, 163, 345

\end{references}
\end{document}